\documentstyle{article}
\renewcommand{\(}{\left (}
\renewcommand{\)}{\right )}
\renewcommand{\=}{\; = \;}
\def\ensuremath#1{\relax\ifmmode#1\else$#1$\fi}
\newcommand{\EV}[1]{\ensuremath{\langle #1 \rangle}}
\newcommand{\k}{\ensuremath{\kappa}}
\newcommand{\ree}{\ensuremath{r_{ee}}}
\newcommand{\EE}{\EV{E}}
\newcommand{\EEC}{\EV{E_C}}
\newcommand{\EEG}{\EV{E_G}}
\newcommand{\EG}{E_G}
\newcommand{\EC}{E_C}
\newcommand{\ereetwo}{\EV{r_{ee}^2}}
\newcommand{\beq}{\begin{equation}}
\newcommand{\eeq}{\end{equation}}
\newcommand{\beqa}{\begin{eqnarray}}
\newcommand{\eeqa}{\end{eqnarray}}
\newcommand{\half}{\frac{1}{2}}
\newcommand{\sqtopi}{\sqrt{\frac{2}{\pi}}}
\newcommand{\sig}{\sigma}
\newcommand{\rsig}{r_{\sig}}
\newcommand{\bsig}{b_{\sig}}
\newcommand{\Lsig}{L_{\sig}}
\newcommand{\sumsig}{\sum_{\sig}}
\newcommand{\sumisig}{\sum_{i\in\sig}}
\newcommand{\sumsigi}{\sum_{\sig\ni i}}
\newcommand{\sumsigij}{\sum_{\sig\ni i,j}}
\newcommand{\sumsigijk}{\sum_{\sig\ni i,j,k}}
\newcommand{\bi}{b_i}
\newcommand{\bj}{b_j}
\newcommand{\bk}{b_k}
\newcommand{\r}{{\bf r}}
\newcommand{\x}{{\bf x}}
\newcommand{\y}{{\bf y}}

\renewcommand{\xi}{\x_i}
\newcommand{\ri}{\r_i}

\newcommand{\rj}{\r_j}
\newcommand{\rvsig}{\r_{\sig}}
\newcommand{\Cppp}{C^{(ppp)}}
\newcommand{\Cptt}{C^{(ptt)}}
\newcommand{\Tr}{\mbox{Tr}\;}
\input{psfig}

\begin{document}
\begin{titlepage}
\begin{flushright}
LU TP 95-25\\
\today \\
\end{flushright}
\vspace{0.5in}
\LARGE
\begin{center}
{\bf Scaling and Scale Breaking in Polyelectrolytes} \\
\vspace{.3in}
\large
Carsten Peterson\footnote{carsten@thep.lu.se},
Ola Sommelius\footnote{ola@thep.lu.se}
and
Bo S\"oderberg\footnote{bs@thep.lu.se}\\
\vspace{0.05in}
Department of Theoretical Physics, University of Lund\\ S\"olvegatan 14A,
S-223 62 Lund, Sweden\\
\vspace{0.25in}

Submitted to {\it Journal of Chemical Physics}

\end{center}
\vspace{0.1in}
\normalsize

Abstract:

We consider the thermodynamics of a uniformly charged polyelectrolyte
with harmonic bonds. For such a system there is at high temperatures
an approximate scaling of global properties like the end-to-end
distance and the interaction energy with the chain-length divided by
the temperature. This scaling is broken at low temperatures by the
ultraviolet divergence of the Coulomb potential. By introducing a
renormalization of the strength of the nearest-neighbour interaction
the scaling is restored, making possible an efficient blocking method
for emulating very large polyelectrolytes using small systems.

The high temperature behaviour is well reproduced by the analytical
high-$T$ expansions even for fairly low temperatures and system
sizes. In addition, results from low-$T$ expansions, where the
coefficients have been computed numerically, are presented. These
results approximate well the corresponding Monte Carlo results at
realistic temperatures.

A corresponding analysis of screened chains is performed. The
situation here is complicated by the appearance of an additional
parameter, the screening length. A window is found in parameter
space, where scaling holds for the end-to-end distance. This window
corresponds to situations where the range of the potential
interpolates between the bond length and the size of the chain. This
scaling behaviour, which is verified by Monte Carlo results, is
consistent with Flory scaling.  Also for the screened chain a blocking
approach can be devised, that performs well for low temperatures,
whereas the low-$T$ expansion is inaccurate at realistic temperatures.

\end{titlepage}

\normalsize

\section{Introduction}

Thermodynamical properties of uniformly charged polyelectrolytes
consisting of linear chains of monomers, with covalent harmonic
bonding forces and Coulomb interactions (screened or unscreened),
have been extensively studied e.g. with Monte Carlo methods
\cite{christos,hooper,higgs,ullner,jon2,stevens} and variational
techniques \cite{jon1,jon2}. Less attention has been paid to the
low-$T$ and high-$T$ expansions for these systems. To some extent the
lack of interest in these two limits has been motivated by the
conjecture that "realistic" temperatures fall outside the range of
such expansions.

In this work we study the scaling properties of polyelectrolytes starting out from
systematic studies of the low-$T$ and high-$T$ expansions both for Coulomb 
and screened chains. The latter are based upon perturbative treatment of 
the interaction. Extensive Monte Carlo (MC) calculations are used to evaluate 
the results.

For the Coulomb chain in dimensionless units there are only two
parameters, the number of monomers $N$ and the temperature $T$. For
global quantities such as the end-to-end distance $\ree$ and the
interaction energy it turns out that one has approximate scaling for
large {\it rescaled temperatures} $\hat{T} = T/N$. This scaling is
broken at low temperatures due to the short-distance divergence of the
Coulomb energy. By appropriate renormalization of the harmonic and
Coulomb terms and by introducing an additional corrective
nearest-neighbor interaction, a blocking scheme results that allows
for fairly accurate MC calculations of very large systems with quite
modest computational investments \cite{pet}. Thus, e.g., when
emulating a $N$=2048 system at room temperature using a system of 50
blocked monomers, approximately a factor $70000$ is gained in computing
speed at the modest expense of $3.6$\% error in $\ree$ \cite{pet}.
%
%
At large $N$ the rescaled temperature becomes small and hence one
should be able to estimate the thermodynamical quantities using a
low-$T$ expansion. Such expansions are systematically developed, and
when evaluating the coefficients numerically one is indeed able to
compute $\ree$ within $3\%$ for large systems at room temperatures.

For a Debye-H\"uckel screened chain the situation is quite
different. In addition to the parameters $N$ and $T$ one has the Debye
screening parameter $\kappa$, introducing an additional length scale
for the chain. For large $\kappa$-values the potential is short range
and the chain is essentially Brownian, whereas for low $\kappa$ the
Coulombic chain is recovered. As $N \rightarrow \infty$ for fixed
$(\kappa, T)$, scaling relations are obtained, consistent with Flory
scaling \cite{flory}. When the range of the potential is much larger
than the average bond length, but much smaller than the overall chain
size, a high temperature expansion indicates a simple dependence on
the combination $N/\kappa^4T^5$ of the deviation from Brownian
behaviour, as measured by $\ree^2/NT$. The blocking approach works
reasonably well also for the screened case, at least for relatively
low temperatures, despite less firm theoretical foundation. The
low-$T$ expansions of global quantities, $\ree$ and the energy, do not
provide as accurate results as in the Coulomb case.

Due to the different nature of the unscreened Coulomb and Debye
screened chains our presentation is structured such, that each of the
two systems is dealt with separately in a self-contained way. We also
include in one of the appendices underlying details of the blocking
approach of ref. \cite{pet}.

The paper is organized as follows: In Section 2 the
Coulomb model is presented together with scaling and scale breaking properties
and numerical results. The corresponding discussion and results for the screened
chain can be found in Section 3. In Section 4 a brief summary is given. Most of
the details concerning high- and low-$T$ expansions are given in the Appendix
A and B respectively.

\newpage

\section{The Coulomb Chain}

\subsection{The Model}

In terms of dimensionless quantities, the energy of a Coulomb chain is
given by \cite{jon2}
\beq
\label{e1}
E^{(N)} \= \EG + \EC \= \half \sum_{i=1}^{N-1} \ri^2 + \sumsig \frac{1}{\rsig}
\eeq
where instead of the absolute monomer positions $\xi$, the {\em bond
vectors}
\beq
\label{ri}
\ri \equiv \x_{i+1}-\xi , \;\; i=1, \ldots, N-1
\eeq
are used, and where $\sig$ runs over contiguous non-nil sub-chains,
with
\beq
\label{rvsig}
\rvsig \equiv \sumisig \ri
\eeq
corresponding to the distance vector between the endpoints of the
subchain.


Due to the simple properties of the Hamiltonian under
rescaling of the coordinates, the thermal expectation values for $\EG$
and $\EC$ are related by a simple virial identity \cite{jon2},
\beq
\label{vir}
	2 \EEG - \EEC \= 3 (N - 1) T
\eeq

\subsection{Zero Temperature}

At zero temperature, the polymer is locked in a minimum energy
configuration: an aligned configuration with the nearest-neighbour
distances $\bi$ satisfying
\beq
\label{ri_0}
	\bi \= \sumsigi \frac{1}{\bsig^2}
\eeq
To a good approximation, this is solved by
\beq
\label{ri_a}
	\bi \approx \left \{ \log \( C \frac{i(N-i)}{N} \) \right \}^{1/3}
\eeq
with $C =\exp(1+\gamma) \approx 4.8$ ($\gamma$ is Euler's constant $\approx 0.5772$).
For $\ree$, $\EG$ and $\EC$, this implies
\beqa
\label{ree_0}
	\ree & \approx & N \( \log N \)^{1/3}
\\ \nonumber
	\EG & \approx & \frac{1}{2} N \( \log N \)^{2/3}
\\ \nonumber
	\EC & \approx & N \( \log N \)^{2/3}
\eeqa
consistently with the virial identity (eq.(\ref{vir})) with $T=0$.

\subsection{Low Temperature Expansions}

At low temperature, one can make an expansion around the minimum energy
configuration. Counting degrees of freedom, removing three for
translational invariance and two for rotational symmetry, we have
\beq
\label{E_low}
	\langle E\rangle \; = \; E_0 + \half \( 3 N - 5 \) T + O\(T^2\)
\eeq
Using this in combination with the virial identity (eq. (\ref{vir})), we obtain for the
partial energies, $\EC$ and $\EG$,
\beqa
\label{ECG_low}
	\EEC & = & \frac{2}{3} E_0 - \frac{2}{3} T + O\(T^2\)
\\ \nonumber
	\EEG & = & \frac{1}{3} E_0 + \( \frac{3}{2} N - \frac{11}{6} \) T + O\(T^2\)
\eeqa
The corresponding expansions for the correlations $\langle \ri \cdot
\rj \rangle$, needed e.g. for $\ree$, can be found in Appendix A,
eq. (\ref{rij_c}). These expansions contain coefficients that have to
be evaluated numerically.

\subsection{High Temperature Expansions}

In the high-$T$ limit, the results for $\ree$ and $\EC$ can be
expanded in power series in $1/T$. For large $N$, the first few terms yield 
(see Appendix B)
\beqa
\label{T_high}
	\ereetwo & \approx & 3 N T + \frac{4}{15} \sqtopi \frac{N^{5/2}}{T^{1/2}}
\\ \nonumber
	\EEC & \approx & \frac{4}{3} \sqtopi \frac{N^{3/2}}{T^{1/2}}
\eeqa
The corresponding expansion for the bond energy $\EG$ can be obtained
from $\EC$ using the virial identity (eq.(\ref{vir})).

We note that the expressions are consistent with a simple scaling 
behaviour since for large $N$ both $\ree/N$ and $\EC/N$ are functions
of the rescaled temperature, $T/N$, only. The relevant small parameter
of the high-$T$ expansion is obviously $N/T$. In Appendix B the
high-$T$ expansion is discussed in some detail.

\subsection{Scaling From a Continuum Formulation}

The scaling properties indicated by the high-$T$ expansion can be
understood as follows. For large $N$, we can approximate the discrete
polymer by a continuous chain, by rescaling $i$, $\xi$ and $T$,
\beq
	\xi = N \y \( \frac{i}{N} \)
,\;\;	T = N \hat{T}
\eeq
in effect replacing the discrete index $i$ by a continuous one, $\tau
= i/N$, $0 < \tau < 1$. Then in terms of $\y(\tau)$, the Boltzmann
exponent $E/T$ is approximated by $\hat{E}/\hat{T}$, with the
{\em effective continuum Hamiltonian} $\hat{E}$ given by
\beq
\label{E_c}
	\hat{E}^{(N)} \; = \; \int_0^1 d\tau \half \dot{\y}(\tau)^2
	+ \int_0^1 d\tau \int_{\tau+1/N}^1 \frac{d\tau'}{|\y(\tau') - \y(\tau)|}
\eeq
The remaining $N$-dependence sits entirely in the short-distance cutoff, necessary to
prevent a potential logarithmic divergence. Without the cutoff, the following naive
scaling behaviour would be exact:
\beq
	\EEC \approx N f(\hat{T})
,\;\;\;	\EEG \approx \frac{3}{2} N^2 \hat{T} + \half N f(\hat{T})
,\;\;\;	\ereetwo \approx N^2 g(\hat{T})
\eeq
Indeed, at high $\hat{T}$ the scale-breaking is not important; the singularity is
washed out by the Brownian fluctuations. From the high-$T$ expansions above, we
infer the asymptotic behaviour of the scaling functions $f$ and $g$:
\beqa
	f(\hat{T}) & \approx & \frac{4}{3} \sqtopi \hat{T}^{-1/2} + O(\hat{T}^{-2})
\\ \nonumber
	g(\hat{T}) & \approx & 3 \hat{T} + \frac{4}{15} \sqtopi \hat{T}^{-1/2} + O(\hat{T}^{-2})
\eeqa
At low $T$, on the other hand, the scaling is more seriously broken:
\beq
	f_N(\hat{T}) \; \approx \; \( \log N \)^{2/3} - \frac{2}{3} \hat{T}
\eeq

\subsection{Blocking Degrees of Freedom}

Neglecting the residual $N$-dependence in eq. (\ref{E_c}), the scaling
properties suggest the possibility of emulating a large system of $N$
monomers by a blocked system of $K=N/Q<N$ effective monomers
\cite{pet}, using the following rescaled Hamiltonian for the latter:
\beq
E_{\mbox{B,naive}}^{(K)}
	= \frac{1}{2Q'} \sum_{i=1}^{K-1} r_{i,i+1}^2
	+ \frac{3}{2}(N-K)T
	\; + \; Q^2 \sum_i \sum_{j>i}\frac{1}{|r_{ij}|}
\label{e1.5}
\eeq
where $r_{ij}$ are distances between blocked monomers, each
representing $Q$ real monomers, while $Q' = (N-1)/(K-1)$. This simple
renormalization technique should work well at high $T$, where the
detailed short-distance properties of the chain are smoothed out by
large fluctuations, and less well at low $T$.

However, the approach can be improved. The $N$-dependence in the
effective continuum Hamiltonian, eq. (\ref{E_c}), implies the
discrepancy
\beq
\label{c2}
	\hat{E}^{(N)} - \hat{E}^{(K)} =
	\int_0^1 d\tau  \int_{\tau+1/N}^{\tau+1/K}\frac{d\tau'}{|\y(\tau')-\y(\tau)|}
\eeq
In the low-$T$ regime, where the chain is more or less linear, the discrepancy is
approximately given by
\beq
\label{c3}
	\hat{E}^{(N)} - \hat{E}^{(K)} \approx
	\log(N/K)\int_0^1\frac{d\tau}{|\dot{\y}(\tau)|}
\eeq
This indicates a possible low-$T$ improvement to the blocking
approach: simply add a corrective interaction term corresponding to
eq.~(\ref{c3}) to the blocked Hamiltonian:
\beqa
\label{e2}
	E_{\mbox{B}}^{(K)} &=& E_{\mbox{B,naive}}^{(K)} + \sum_{i=1}^K W(r_{i,i+1})
\\ 
\label{W}
	W(r) &=& \frac{Q^2\log(Q)}{r}
\eeqa
With this improved blocked Hamiltonian, very large chains can be
emulated (using MC) with relatively small blocked systems with quite
small errors all the way down to $T=0$ \cite{pet}.

The correction term can also be derived directly for the discrete
system by considering eq. (\ref{e2}), with an unspecified corrective
nearest-neighbour interaction $W(r)$.
At low temperatures we approximate the nearest-neighbour distance by a
constant $a$. For the blocked system, the corresponding
nearest-neighbour distance then is $b=Qa$. In this approximation, the
Coulomb energy for large $N$ is given by
\beq
	U_N \approx \frac{N}{a} \sum_l^N \frac{1}{l}
	\approx \frac{N}{a} \log N
\eeq
while the blocked interaction energy for large $K$ reads
\beq
	U_K \approx \frac{Q^2K}{b} \log K + K W(b)
\eeq
Choosing $W(r)$ such that $U_N = U_K$ for $b=Qa$ yields precisely
eq. (\ref{W}).

Appendix B contains a discussion of the high-$T$ properties of the
blocked system.

\subsection{Numerical Comparisons}

In this section we compare the results from the high- and low-$T$ expansions
as well as scaling properties with MC data. All MC calculations, both
for Coulomb and screened chains (see next section) were performed using the
pivot algorithm \cite{pivot}. Most results are based on $10^4$ thermalization
sweeps and $10^5$ measured configurations. Some of the MC results were taken
from ref. \cite{jon2}. For results from exploring the blocking method we
refer the reader to ref. \cite{pet}.

\subsubsection{Low T Results}

\begin{figure}[htb]
\centering
\mbox{
	\psfig{figure=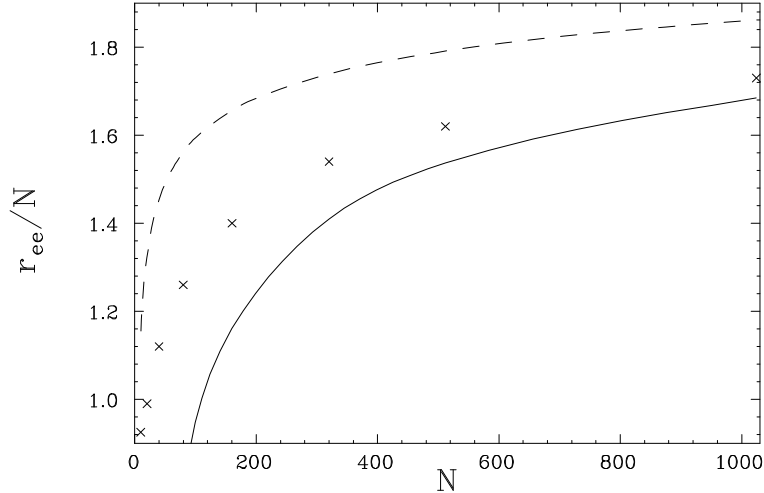,width=10cm}
}
\caption{$\ree/N$ as a function of $N$ for a Coulomb chain at $T$=$T_r$.
Dashed line, full line and crosses represent $T$=0 (eq. (\protect\ref{ree_0})),
first order low-$T$ expansion (eq. (\protect\ref{rij_c})) and MC results,
respectively.}
\label{ree_low_T}
\end{figure}

It is of interest to investigate what low $T$ means in terms of
physical temperatures. In the dimensionless units of eq. (\ref{e1}),
room temperature (290K) corresponds to $T_r$=0.837808. We have
computed $r_{ee}/N$ , where $\ree$=$\sqrt{\ereetwo}$, using the
low-$T$ expansion corresponding to eq. (\ref{rij_c}) for different
chain sizes $N$ at $T=T_r$.  In fig. \ref{ree_low_T}, comparisons are
made between these low-$T$ results and MC data. Also shown in
fig. \ref{ree_low_T} is the $T$=0 solution (cf. eq. (\ref{ree_0})).
Not surprisingly the low-$T$ approximation becomes more accurate as
$N$ increases since it corresponds to a lower $\hat{T}$. For $N$=1024
the error is less than 3\%, which is quite impressive.  Computing
$r_{ee}$ for $N$=1024 using eq. (\ref{rij_c}) is $\sim 200$ times
faster than using MC.  The computational demand for both methods grows
like $N^3$.

\subsubsection{Scaling}

\begin{figure}[htb]
\centering
\mbox{
	\psfig{figure=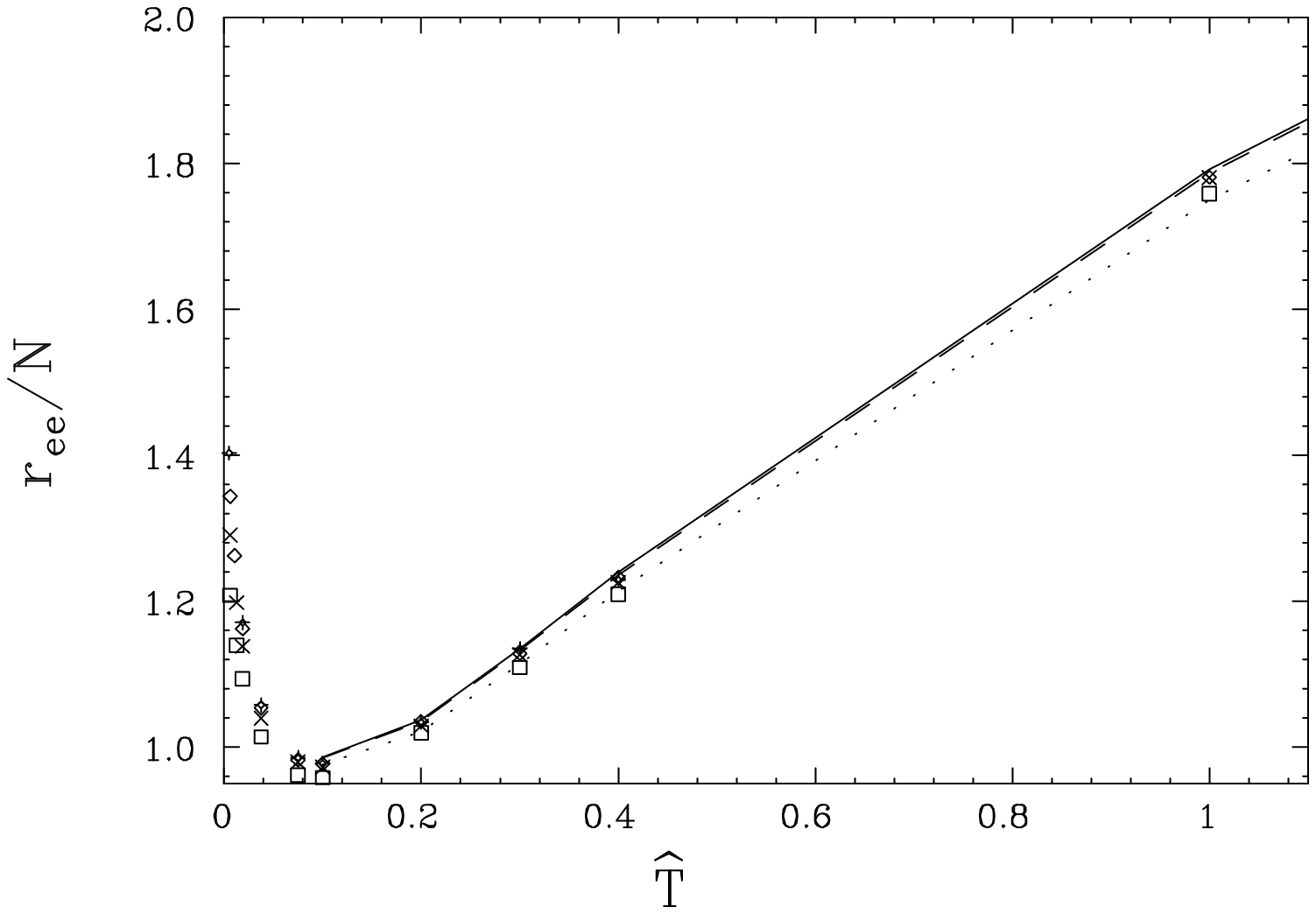,width=9.7cm}
}
\caption{$\ree/N$ as a function of $\hat{T}$ for a Coulomb chain from MC
results for $N$=20 ($\Box$), 40 ($\times$), 80 ($\diamond$) and 160 (+).
The lines show second order high-$T$ expansion results for $N$=20 (dotted),
 80 (dashed) and $\infty$ (full)
from eqs. (\protect\ref{T_high_N},\protect\ref{T_high_inf}).}
\label{ree_scale}
\end{figure}

\begin{figure}[htb]
\centering
\mbox{
	\psfig{figure=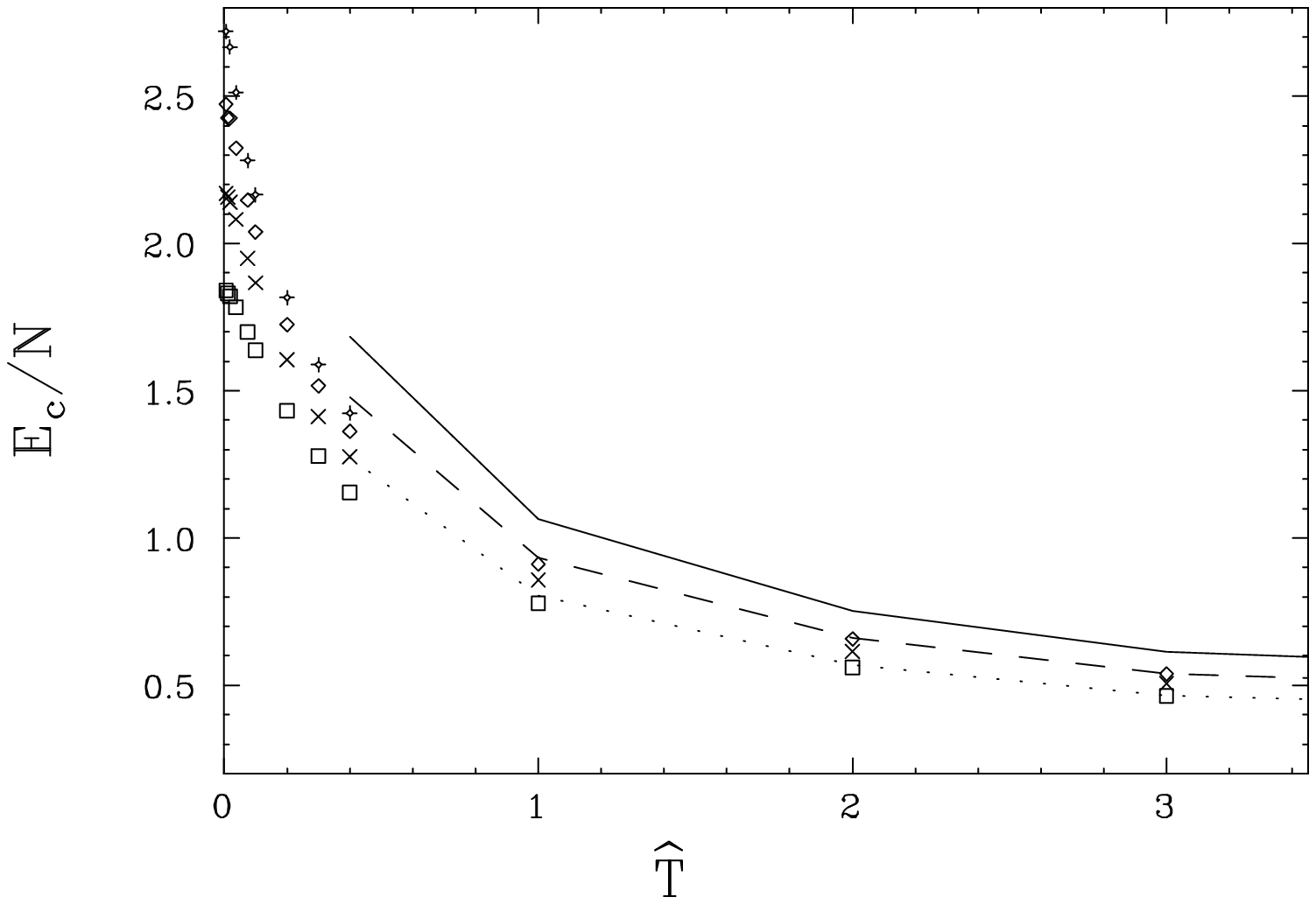,width=9.7cm}
}
\caption{$\EC/N$ as a function of $\hat{T}$ for a Coulomb chain from MC
results for $N$=20 ($\Box$), 40 ($\times$), 80 ($\diamond$) and 160 (+).
The lines show leading order high-$T$ results for $N$=20 (dotted), 80 (dashed)
and $\infty$ (full) from eqs. (\protect\ref{T_high_N},\protect\ref{T_high_inf}).}
\label{EC_scale}
\end{figure}

Next we test the scaling properties expected from the high temperature
expansions (eq. (\ref{T_high}) and Appendix B) against numerical
results. In figs. \ref{ree_scale} and \ref{EC_scale}, $\ree/N$ and
$\EC/N$ are shown as functions of $\hat{T}=N/T$. The predictions from
the high-$T$ expansion for $\ree$ fit extremely well for
$\hat{T}>0.2$, whereas for $\EC$ the validity region of this expansion
is located at much larger $\hat{T}$. Also shown are the finite-$N$
effects from eqs. (\ref{T_high_N},\ref{T_high_inf}), which are larger
for $\EC$ than for $\ree$ as can be understood from
eqs.~(\ref{B2},\ref{T_high_N},\ref{T_high_inf}). The deviation from
scaling in fig. \ref{EC_scale} for $\hat{T}>1$ is almost entirely due
to finite-$N$ effects.

\section{Debye Screened Chains}

Next we turn to the case of the Coulomb chain being screened by the
surrounding medium. We model this effect by using the Debye-H\"uckel
potential (being fully aware of its limitations).

\subsection{The Model}

The energy of the Debye screened chain takes the following form \cite{jon2}:
\beq
	E(\r) = \EG + \EC = \half \sum_{i=1}^{N-1} \ri^2 +
	\sumsig \frac{e^{-\kappa \rsig}}{\rsig}
\eeq
where the additional parameter $\kappa$ is the inverse screening length. In what
follows we assume that $\kappa$ is independent of $T$.


In the screened case, a proper virial identity relating the average bond and
interaction energies does not exist (cf. eq. (\ref{vir})). Still, one has the
slightly less useful identity \cite{jon2}.
\beq
\label{vir1}
	2 \langle \EG \rangle - \langle \EC \rangle - \kappa
	\left \langle \sumsig e^{-\kappa \rsig} \right \rangle
	= 3(N-1)T
\eeq

\subsection{Zero Temperature}

Since the screened interaction is short-range, the ground-state configuration
can, for large enough $N$, be considered translation-invariant, with a
constant, $N$-independent nearest-neighbour distance $a$. Defining $\eta
\equiv \kappa a$, we have
\beq
	a^3 = \frac{\eta}{e^\eta-1} - \log(1-e^{-\eta})
\eeq
valid for $N\eta \gg 1$. This gives $a$ implicitly as a function of $\kappa$,
and we have, for small $\kappa$:
\beqa
\nonumber
	a & \approx & (-\log \eta)^{1/3}
\\ \nonumber
	\kappa & \approx & \frac{\eta}{(-\log \eta)^{1/3}}
\\
	\Rightarrow a & \approx & \left ( - \log \kappa \right )^{1/3}
\eeqa
while for large $\kappa$, we obtain
\beqa
\nonumber
	a & \approx & \left ( (\eta + 1) e^{-\eta} \right )^{1/3}
\\ \nonumber
	\kappa & \approx & \frac{\eta}{\left ( (\eta + 1) e^{-\eta} \right )^{1/3}}
\\
	\Rightarrow a & \approx & 3 \frac{\log \kappa}{\kappa}
\eeqa
Similarly, we have for the energy per monomer
\beqa
	E_G/N & = & \frac{a^2}{2}
\\
	E_C/N & = & -\frac{1}{a}\log\left(1-e^{-\eta}\right)
\eeqa

\subsection{Low Temperature Expansions}

Also in the screened case, observables can be expanded in a low temperature
series, analogous to the Coulomb case, although the details are slightly modified
(cf. Appendix A). The expression for $\EE$, eq.~(\ref{E_low}), depends only
on the number of degrees of freedom and is still valid, while the expressions for
$\EEC$ and $\EEG$ separately (eqs.~\ref{sEc}, \ref{sEg}) are less transparent
than the corresponding unscreened equations (\ref{ECG_low}).

\subsection{High Temperature Expansions}

Assuming a $T$-independent $\kappa$, and a large $N$, we have in the
high-$T$ limit
\beqa
	\langle r_{ee}^2 \rangle & \approx &
	3 (N-1) T + 4 \sqtopi N^{3/2} \kappa^{-2} T^{-3/2}
\label{highscT}
\\
	\EEC & \approx &
	\zeta(\frac{3}{2}) \sqtopi N \kappa^{-2} T^{-3/2}
\label{EChighscT}
\eeqa
while $\EEG$ in this limit is given by $\frac{3}{2} (N-1) T +
\frac{3}{2} \EEC$. In this case, there are no obvious scaling
properties.

A Flory estimate of the free energy for the Debye-H\"{u}ckel potential
gives a radius of the chain
\beq
	r \sim \frac{N^{3/5}}{\kappa ^ {2/5}}
\label{flory}
\eeq
indicating that the chain will behave like a self-avoiding walk in the large-$N$
limit. Both Flory scaling and the high temperature expansion,
eq. (\ref{highscT}), are consistent with a more general scaling behaviour:
\beq
	\frac{r^2}{NT} \sim f\left( \frac{N}{\kappa^4 T^5} \right)
\label{scale}
\eeq
where the function $f(x)$ behaves as $3 + 3.2\, x^{1/2}$ for small $x$,
while an asymptotic behaviour $f(x) \sim x^{1/5}$ would correspond to Flory scaling.

\subsubsection*{Scaling window}

A reasonable assumption is that the relevant scaling region is where
the range of the potential is small compared to the chain size, and
large compared to the monomer-monomer separation, i.e.
\beq
\label{k2nt}
1 \ll \kappa^2 N T \ll N
\eeq
In this window, the leading expression for $\EEC$, eq. (\ref{EChighscT}),
has to be modified into
\beq
	\EEC \approx 2 N \kappa^{-1} T^{-1}
\eeq
with $\EEG = \frac{3}{2}(N-1)T + \EEC$, while the expression for
$\ereetwo$, eq. (\ref{highscT}), is unaffected.

\subsection{Blocking Degrees of Freedom}

Despite the fact that the linear chain approximation is of more limited validity
in the screened case, one still can devise a blocking approach. Indeed, for low enough 
temperatures the approach works quite well \cite{pet}.

In this case the discrete derivation of the correction term is more convenient. Again,
starting  out from a discrete blocked energy
\beq
E_{\mbox{B}}^{(K)}
	= \frac{1}{2Q} \sum_{i=1}^{K-1} r_{i,i+1}^2
	+ \frac{3}{2}(N-K)T
	\; + \; Q^2 \sum_i \sum_{j>i} \frac{e^{-\kappa r_{ij}}}{r_{ij}}
	+ \sum_{i=1}^K W(r_{i,i+1})
\label{e2_s}
\eeq
following the same steps as in the unscreened case, one arrives at
%
%
%
%
%
%
\beq
	W(r)= -\frac{Q^2}{r} \left [
	\log \left ( 1-e^{-\kappa r / Q} \right )
	- \log \left ( 1-e^{-\kappa r} \right ) \right ]
\label{Ws}
\eeq
In this way, the longitudinal behaviour will be rather accurately
reproduced by the blocked system. However, at low $T$, the transverse
fluctuations will dominate, as indicated by eq. (\ref{V2}).
These are probably more difficult to reproduce than the longitudinal
fluctuations.

\subsection{Numerical Comparisons}

In this section we investigate the validity regime of the high- and
low-temperature expansions of $r_{ee}$ for the screened chain. In
addition, we test the approximate scaling indicated by
eq. (\ref{scale}), and the corresponding scale-breaking arising
outside the scaling window given by eq. (\ref{k2nt}).

\begin{figure}[htb]
\centering
\mbox{
	\psfig{figure=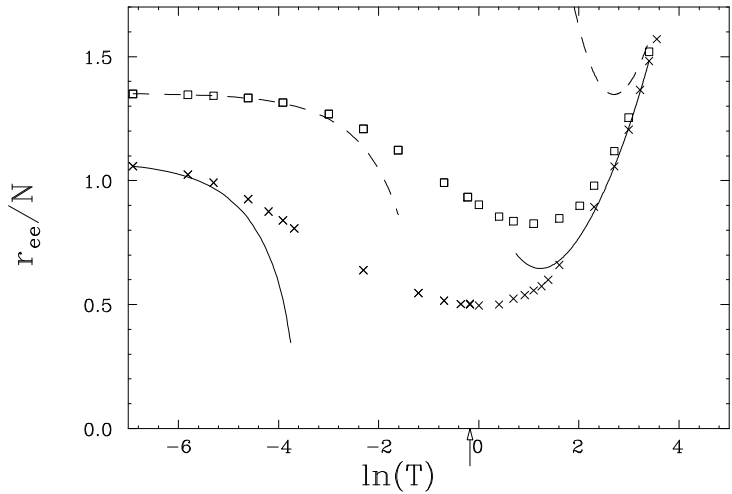,width=9.7cm}
}
\caption{$\ree/N$ as a function of $T$ for a screened, $N$=40, Coulomb chain
with
$\kappa = 0.1 $($\Box$) and 0.63 ($\times$).
 The lines show second order high-$T$ expansion results ($\kappa$=0.1 (dashed),
 0.63 (solid)) and the corresponding first order low-$T$ expansions.
$T_r$ is marked with an arrow.}
\label{ree_screened}
\end{figure}

In fig. \ref{ree_screened} the high- and low-$T$ expansions of the
end-to-end distance, eqs. (\ref{highscT}, \ref{rij_s}), are compared
to MC data.  Clearly, an increase in $\kappa$ enlarges the high-$T$
interval and decreases the low-$T$ interval. Quantities at room
temperature $T_r$, for biologically relevant values of $\kappa$, are
hard to estimate from both high- and low-$T$ expansions.

\begin{figure}[htb]
\centering
\mbox{
	\psfig{figure=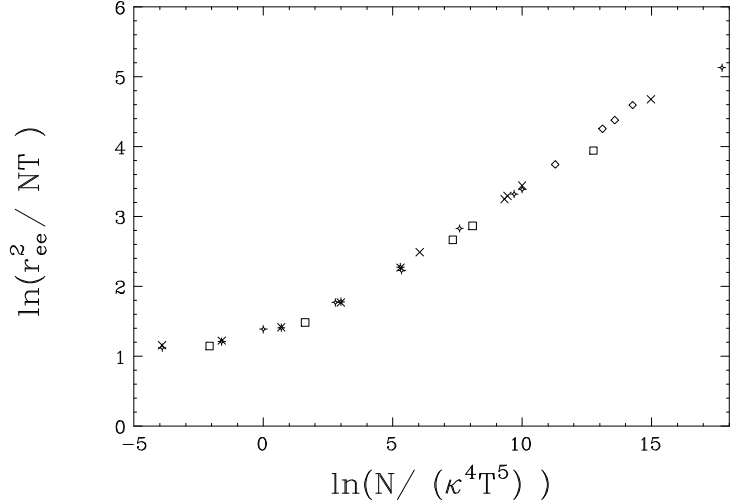,width=9.7cm}
}
\caption{$\ln r_{ee}^2/NT$ as a function of $\ln N/ \kappa^4 T^5 $
with $N\kappa^2 T \in (8, 0.3 N)$.}
\label{r2oNT}
\end{figure}

\begin{figure}[htb]
\centering
\mbox{
	\psfig{figure=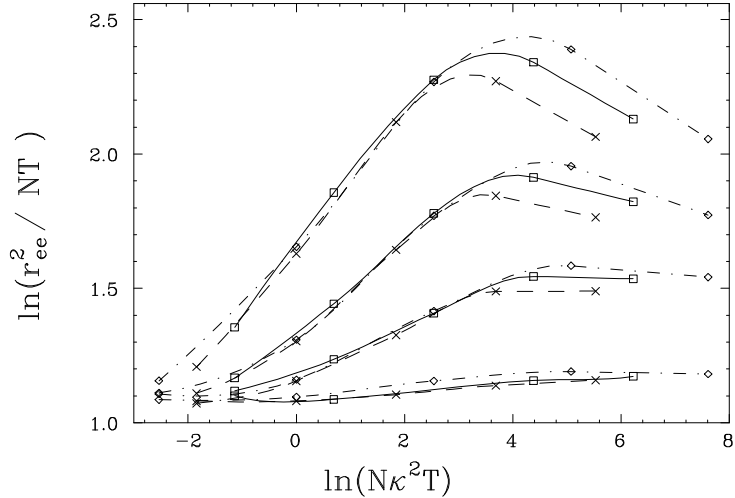,width=9.7cm}
}
\caption{$\ln (r_{ee}^2/NT)$ as a function of $\ln(N \kappa^2 T)$.
The four collection of curves corresponds to four different values of $\ln N/(\kappa^4T^5)$:
 from top to bottom $5.30$, $2.99$, $0.69$ and $-3.91$ for $N=40$$(\Box)$,
 $80$$(\times)$ and $160$($\Diamond$). The lines are only intended
to guide the eye.}
\label{scalebr}
\end{figure}

The approximate scaling as suggested by high temperature expansions
and supported by Flory arguments (eq. (\ref{scale})) is shown in
fig. \ref{r2oNT} to hold almost too good for the somewhat arbitrary
cut $N\kappa^2 T \in (8, 0.3 N)$. In fig. \ref{scalebr} the deviation
from this behaviour is shown by plotting $\ln (r_{ee}^2/NT)$ as a
function of $\ln N\kappa^2T$ for fixed values of $N /
(\kappa^4T^5)$. Small values of $N\kappa^2T$ then are equivalent to a
high temperature Coulombic limit, while large values ($\gg N$) correspond to a
low temperature Gaussian limit.
Hence in both
limits the curve approaches the Gaussian value, which is $\ln{3}$.

A noticeable feature of fig. \ref{scalebr} is the seemingly
$N$-independent rise to the left, indicating that in this region
$\ereetwo$ depends only on $N/\kappa^4T^5$ and $N\kappa^2T$, or,
equivalently, on $\hat{T}=T/N$ and $\hat{\kappa}=N\kappa$. This is
equivalent to stating that a naive (i.e. without the extra
nearest-neighbour term) blocking approach should work well there. The
scale-breaking seen in the right part of fig. \ref{scalebr}, where the
curves bend downwards, occurs when $N\kappa^2T$ becomes comparable to
$N$, violating the assumed scaling condition, eq. (\ref{k2nt}). For an
infinite $N$, this scale-breaking would be absent, and the
corresponding curve would never bend down.
%


\section{Summary}

Scaling and scale-breaking properties of thermodynamic and
conformational properties for simple models of charged polymer chains
have been established.  These properties are deduced from high-and
low-$T$ expansions by means of a perturbative treatment of the
interaction for both Coulomb and Debye-H\"{u}ckel screened chains. In
addition to the naive high-$T$ limit with fixed $\kappa$ for the
screened chains we also consider a restricted limit where $T \to
\infty$ and $\kappa \to 0$ while keeping $N\kappa^2T$ large and
$\kappa^2T$ small.
 
For the Coulomb chain, an increase in $N$ makes the chain stiffer due
to the long-range character of the electrostatic interaction. The same
effect is obtained by lowering the temperature; indeed, thermodynamic
quantities for long chains are well approximated by a low-temperature
expansion. The relevant variable of the chain is the rescaled
temperature $\hat{T} = T/N$, as further supported by high-$T$
expansions which for large $N$ become equivalent to high-$\hat{T}$
expansions.

For the screened chain the situation is somewhat different due to the
appearance of a new length scale -- the inverse screening length
$\kappa^{-1}$. For $\kappa^{-1}$-values in a window between the bond
length and the chain length, a relevant scaling variable $N / (
\kappa^4 T^5)$ can be identified. The associated scaling behaviour is
consistent with Flory scaling \cite{flory}.

The scaling properties in terms of $\hat{T} = T/N$ provides a firm
theoretical basis for the real space blocking scheme of
ref. \cite{pet}; a detailed derivation is
given. As reported in ref. \cite{pet} this approach is
very powerful for computing $\ree$ and $\EC$, both for Coulomb and
screened chains, with low errors.  Nothing prohibits the extension of
the blocking scheme to include titration and multi-chain systems.

For relatively small temperatures, the low-$T$ expansion from
ref. \cite{jon2} is further developed into an efficient computational
tool for computing conformational properties. In the Coulomb case
results at room temperatures with low errors emerge from this
scheme. For the screened case the low-$T$ approach is not viable, at
least for reasonable values of $N$, $T$ and $\kappa$.

%

\newpage

%
\setcounter{section}{0}
\renewcommand{\thesection}{Appendix \Alph{section}.}
\renewcommand{\theequation}{\Alph{section}\arabic{equation}}

\section{Low T Expansions}
\setcounter{equation}{0}

\subsection*{Coulomb Chains}

This section has a certain unavoidable overlap with Appendix D in
ref. \cite{jon2}. Low-$T$ expansions for $\EEC$ and $\EEG$ are given in
eq.  (\ref{ECG_low}). Computing low-$T$ corrections to $\ree$ is
somewhat more elaborate.

In terms of the zero temperature bond lengths defined by $\bi \equiv
\sumsigi \frac{1}{b_{\sig}^2}$ we define a set of tensors
\beqa
	B_{ij} & = & \sumsigij\frac{1}{b_{\sig}^3}
\\	C_{ijk} & = & \sumsigijk \frac{1}{b_{\sig}^4}
\eeqa
In addition, we need the two matrices related to $B$,
\beqa
	U & = & (1 + 2B)^{-1}
\\	V & = & P(1 - B)^{-1}P \label{V1}
\eeqa
where $P$ denotes the projection matrix onto the subspace orthogonal
to $b$, which is a zero-mode of $1-B$.

In terms of these tensors, we have the quadratic expectation-values at
low-$T$:
%
\beq
\label{rij_c}
	\langle \ri \cdot \rj \rangle = \bi \bj + T \left ( U_{ij} + 2 V_{ij} +
	\frac{4 \bi \bj}{3\sum_k \bk^2} + 3 \sum_{klm} C_{klm} (\bi U_{jk} + \bj
	U_{ik}) ( U_{lm} - V_{lm} ) \right ) + O(T^2)
\eeq
where the first two terms of the $T$ coefficient are the naive
contributions from the longitudinal and transverse fluctuations. The
rest are corrections due to the rotational degeneracy of the $T=0$
configuration, which is also responsible for the transverse zero-modes
(of $1-B$). In terms of this expression, $\langle r_{ee}^2 \rangle$ is
obtained by summing independently over $i$ and $j$, while $\langle E_G
\rangle$ is obtained by summing over $i=j$ and dividing by two.

\subsection*{Screened Chains}

For a screened chain, with $\bi$ still denoting the (now $\kappa$ dependent)
zero temperature bond lengths, the tensors take the following form
\beqa
	B & = &{\sumsigij} \frac{ e^{-\kappa \bsig} }{b^3} \left( 1 + \kappa \bsig \right)
\\
	U & = &\left( 1 + \kappa^2 \sumsigij \frac{ e^{-\kappa b}}{\bsig} + 2 B \right)^{-1}
\\
	V & = &P\left( 1 - B \right)^{-1}P    \label{V2}
\eeqa
while two versions of the $C$-tensor are needed, one related only to parallel
fluctuations $\Cppp$ and the other, $\Cptt$, coupled to one parallel and two
transverse fluctuations.
\beqa
	\Cppp_{ijk} & = & {\sumsigijk} \frac{ e^{-\kappa b}}{b^4}
	\left( 1 + \kappa b + \frac{\kappa^2 b^2}{3} \right)
\\
	\Cptt_{ijk} & = & {\sumsigijk} \frac{ e^{-\kappa b}}{b^4}
	\left( 1 + \kappa b + \frac{\kappa^2 b^2}{2} +
	\frac{\kappa^3 b^3}{6} \right)
\eeqa
In terms of these tensors we have, to first order in the low-$T$ expansion,
the expectation values
\beq
\label{rij_s}
\langle \ri \cdot \rj \rangle = \bi \bj + T \left ( U_{ij} + 2 V_{ij} +
2 \frac{ \sum_k \bk \left( \bj U_{ik} + \bi U_{jk} \right)}{\sum \bk^2}
 + 3 \sum_{klm} \left( \Cppp_{klm} U_{lm} - \Cptt_{klm} V_{lm} \right)
 \left( \bj U_{ik} + \bi U_{jk} \right)
 \right)
\eeq
\beqa
\label{sEc}
	\langle E_C \rangle & = &\sumsig \frac{e^{-\kappa b}}{b}
	- T \sum_i \left( 2 \frac{ \sum_j \bi \bj U_{ij}}{\sum \bk^2}
	+ 3 \sum_{klm} \left( \Cppp_{klm} U_{lm} - \Cptt_{klm} V_{lm} \right)
	\bi U_{ik} \right)
\nonumber \\
	& & - \frac{T}{2} \left( \Tr U + 2 \Tr V \right) + \frac{3N - 5}{2} T
\\
\label{sEg}
	\langle E_G \rangle & = & \frac{1}{2}\sum_i \bi^2 + T \sum_i\left(
	2 \frac{ \sum_j \bi \bj U_{ij}}{\sum \bk^2}
	+ 3 \sum_{klm} \left( \Cppp_{klm} U_{lm} - \Cptt_{klm} V_{lm} \right)
	\bi U_{ik} \right)
\nonumber \\
	& & + \frac{T}{2} \left( \Tr U + 2 \Tr V \right)
\eeqa
%

\newpage
\section{High $T$ Expansions}
\setcounter{equation}{0}

\subsection*{Sums at large $N$}

In what follows, we will encounter sums over subchains like the following for large
$N$
\beq
	S_{\alpha} \equiv \sumsig \Lsig^{\alpha}
	\equiv \sum_{k=1}^N (N-k) k^{\alpha}
\eeq
where $\Lsig$ is the length (i.e. the number of bonds) of the subchain
$\sigma$.  Obviously, there exist $N-k$ distinct subchains of length $k$.

In particular, we will encounter the following sums:
\beqa
\nonumber
	S_{1/2} & = & \frac{4}{15} N^{5/2}
	+ B N - \frac{1}{12} N^{1/2} - A + \ldots
\\
\label{B2}
	S_{-1/2} & = & \frac{4}{3} N^{3/2}
	+ C N - B - \frac{1}{12} N^{-1/2} + \ldots
\\ \nonumber
	S_{-3/2} & = & D N - 4 N^{1/2} - C
	- \frac{1}{12} N^{-3/2} + \ldots
\eeqa
where the numerical values of the constants $A$, $B$, etc. are
\beq
	A \approx -.02548 ,\;\;
	B \approx -.20789 ,\;\;
	C \approx -1.46035 ,\;\;
	D \equiv \zeta(3/2) \approx 2.612375
\eeq
In addition, we need
\beq
	\zeta(3) \approx 1.202057
\eeq

\subsection*{High $T$ expansions for a Coulomb chain}

%
%
%
%
For a system of $N$ monomers, we have the following high-$T$ expansions,
which are obtained by considering the Coulomb interaction as a perturbation,
\beqa
\label{T_high_N}
\ereetwo & \approx &
	3(N-1)T + \sqtopi T^{-1/2} \sumsig \Lsig^{1/2}
\\ \nonumber
\EEG & \approx &
	\frac{3}{2}(N-1)T + \frac{1}{2} \sqtopi T^{-1/2} \sumsig \Lsig^{-1/2}
\\ \nonumber
\EEC & \approx &
	\sqtopi T^{-1/2} \sumsig \Lsig^{-1/2}
\eeqa
At large $N$ these simplify to
\beqa
\label{T_high_inf}
\ereetwo & \approx &
	3 (N-1) T + \frac{4}{15} \sqtopi T^{-1/2} N^{5/2}
\\ \nonumber
\EEG & \approx &
	\frac{3}{2} (N-1) T + \frac{2}{3} \sqtopi T^{-1/2} N^{3/2}
\\ \nonumber
\EEC & \approx &
	\frac{4}{3} \sqtopi T^{-1/2} N^{3/2}
\eeqa
The corresponding high-$T$ expansions for a blocked system ($N
\rightarrow K = N/Q$) yield identical expressions to the order shown,
as long as $K$ is large. This can be understood by considering the
blocked Hamiltonian of eq. (\ref{e2}), with $E_G$ defined to include
the constant $\frac{3}{2}(N-K)T$, and $E_C$ to include the corrective
nearest-neighbour interaction. The latter term, given by
eq. (\ref{W}), does not contribute in the high-$T$ limit: it is needed
only at low $T$.
%
%
%

\subsection*{Generic Perturbation Expansion for Screened Chain}

For a Debye screened chain, one must distinguish between two distinct
high-$T$ limits -- (1) the naive one, where $T \to \infty$ with a
constant \k, and (2) a modified limit, corresponding to the scaling
window, eq. (\ref{scale}), where $T \to \infty$ and $\kappa \to 0$
keeping $\k^2T$ small and $N\k^2T$ large. The latter of course
requires a large $N$.

In both cases the analysis is based on a perturbative treatment of the
interaction, which to a few orders yields the following results for a
screened chain of $N$ monomers.
\beqa
\ereetwo & \approx &
	3(N-1)T + \sqtopi \kappa^3 T \sumsig \Lsig^2 F_2(\kappa\sqrt{\Lsig T})
\\ \nonumber
\EEG & \approx &
	\frac{3}{2}(N-1)T + \frac{1}{2} \sqtopi \kappa^3 T \sumsig \Lsig F_2(\kappa\sqrt{\Lsig T})
\\ \nonumber
\EEC & \approx &
	\sqtopi \kappa \sumsig F_1(\kappa\sqrt{\Lsig T})
\eeqa
where the $F_k()$ are erf-related functions, defined by
\beq
	F_k(x) = (2k-1)!! \exp(x^2/2) \int_x^{\infty} dy\; y^{-2k}\exp(-y^2/2)
\eeq

\subsection*{Conventional High $T$ expansion for a Screened Chain}

In the $T\to \infty$ limit with a constant \k\ and a fixed large $N$,
the sum in the expression for \ereetwo\ is dominated by the large $L$
terms (scaling) and can be approximated by an integral. In contrast,
the sums in \EEC\ and \EEG\ are dominated by small $L$ (no scaling),
and must be explicitly summed.  We thus obtain
\beqa
\ereetwo & \approx &
	3 (N-1) T + 4 \sqtopi N^{3/2} \kappa^{-2} T^{-3/2}
\\ \nonumber
\EEG & \approx &
	\frac{3}{2} (N-1) T + \frac{3}{2} \EEC
\\ \nonumber
\EEC & \approx &
	\zeta(\frac{3}{2}) \sqtopi N \kappa^{-2} T^{-3/2}
\eeqa

The corresponding high-$T$ expansions for a blocked system, based on
eqs. (\ref{e2_s}, \ref{Ws}) with large $K$ and large $Q$, yield for
\ereetwo\ the same as above. However, \EEG\ and \EEC\ become too large,
\beqa
\EEG & \approx &
	\frac{3}{2}(N-1)T + \frac{3}{2} \EEC
\\ \nonumber
\EEC & \approx &
	\sqtopi \zeta \( \frac{3}{2} \) N \kappa^{-2} T^{-3/2}
	\left (
	Q^{-1/2} + \frac{\zeta(3)}{\zeta(\frac{3}{2})} Q^{3/2}
	\right )
\eeqa
This is due to an overshooting by the corrective term, eq. (\ref{Ws}),
which gives rise to the $Q^{3/2}$ term in the expression for \EEC .
On the other hand, upon omitting the corrective term, \EEC\ becomes
too small by a factor $Q^{-1/2}$, due to the lack of scaling mentioned
above.  Obviously, for thermodynamic properties, the blocking method
does not apply in this limit.

\subsection*{Modified High $T$ expansion for Screened Coulomb}

With the restriction of staying within the scaling window, $\k^2T \ll 1 \ll N\k^2T$,
all three sums scale, i.e. they are dominated by the large-$L$ terms. While 
the expression for \ereetwo\ remains unchanged, we obtain for \EEG\ and \EEC\ 
in this limit
%
%
%
\beqa
\EEG & \approx &
	\frac{3}{2} (N-1) T + \EEC
\\ \nonumber
\EEC & \approx &
	2N\k^{-1}T^{-1}
\eeqa

For a system blocked into $K$ effective monomers, these expressions
are reproduced, provided also the blocked system is in the scaling
window, i.e. if $\k^2TQ \ll 1$; otherwise it yields to too small
values for \EEC\ and \EEG.
%



\newpage
\begin{thebibliography}{99}

\bibitem{christos} G. A. Christos and S. L. Carnie,
{\it J. Chem. Phys.} {\bf 91}, 439 (1089).

\bibitem{hooper} H.H. Hooper and H.W. Blanch and J. M. Prausnitz,
{\it Macromolecules} {\bf 23}, 4820 (1990).

\bibitem{higgs} P.G. Higgs and H. Orland
{\it J. Chem. Phys.} {\bf 95}, 4506 (1991).

\bibitem{ullner} M. Ullner and B. J\"{o}nsson and P.-O. Widmark,
{\it J. Chem. Phys.}{\bf 100}, 3365 (1994).

\bibitem{jon2} B. J\"onsson, C. Peterson and B. S\"oderberg,
{\it J. Phys. Chem.} {\bf 99}, 1251 (1995).

\bibitem{stevens} M.J. Stevens and K. Kremer,
{\it J. Chem. Phys.} {\bf 1669}, 4 (1995).

\bibitem{jon1}B. J\"onsson, C. Peterson and B. S\"oderberg,
{\it Phys. Rev. Lett.} {\bf 71}, 376 (1994).

\bibitem{pet} C. Peterson, O. Sommelius and B. S\"oderberg,
{\it Phys. Rev. Lett.} {\bf 76}, 1079 (1996).


\bibitem{flory} P.J. Flory,
{\it J. Chem. Phys.} {\bf 17}, 303 (1949).

\bibitem{pivot} M. Lal, {\it Mol. Phys.} {\bf 17}, 57 (1969);\\
N. Madras and A.D. Sokal, {\it J. Stat. Phys.} {\bf 50}, 109 (1988).
{\bf 50}, 109 (1988).

\bibitem{irback} A. Irb\"ack, {\it J. Chem. Phys.} {\bf 101}, 1661 (1994).

\end{thebibliography}
\end{document}